\documentstyle[aps,prl,twocolumn]{revtex}

\begin{document}
\flushbottom
\twocolumn[\hsize\textwidth\columnwidth\hsize\csname @twocolumnfalse\endcsname
\draft

\title{Linear-scaling ab-initio calculations for large and complex systems}

\author{Emilio Artacho$^{1*}$, Daniel S\'anchez-Portal,$^2$ 
Pablo Ordej\'on,$^3$ Alberto Garc\'{\i}a,$^4$ and Jos\'e M. Soler$^5$}

\address{
$^1$P\^ole Scientifique de Mod\'elisation Num\'erique,
Ecole Normale Sup\'erieure de Lyon, 69364 Lyon, Cedex 07, France \\
$^2$Department of Physics and Materials Research Laboratory, University of 
Illinois, Urbana, Illinois 61801, USA \\
$^3$Departamento de F\'{\i}sica, Universidad de Oviedo, 33007 Oviedo, Spain \\
$^4$Departmento de Fisica Aplicada II, Universidad del Pais Vasco, 
48080 Bilbao, Spain \\
$^5$Depto. de F\'{\i}sica de la Materia Condensada, C-III,
and Inst. Nicol\'as Cabrera, Universidad Aut\'onoma, 28049 Madrid, Spain}

\date{\today}

\maketitle

\begin{abstract}
A brief review of the {\sc Siesta} project is presented in the context of
linear-scaling density-functional methods for electronic-structure
calculations and molecular-dynamics simulations of systems with a large 
number of atoms. Applications of the method to different systems are
reviewed, including carbon nanotubes, gold nanostructures, adsorbates on
silicon surfaces, and nucleic acids. Also, progress in atomic-orbital
bases adapted to linear-scaling methodology is presented.
\end{abstract}

\pacs{ }

]

\section{INTRODUCTION}

It is clearer every day the contribution that first-principles calculations
are making to several fields in physics, chemistry, and recently 
geology and biology. The steady increase in computer power and the progress 
in methodology have allowed the study of increasingly more complex and 
larger systems \cite{rmp}. It has been only recently that the scaling 
of the computation expense with the system size has become an important issue
in the field. Even efficient methods, like those based on 
Density-Functional theory (DFT), scale like $N^{2\mbox{-}3}$, being $N$ the 
number of atoms in the simulation cell. \cite{rmp} This problem stimulated the 
the first ideas for methods which scale linearly with system size
\cite{Ordejon95}, a field that has been the subject of important efforts 
ever since \cite{review}.

The key for achieving linear scaling is the explicit use of locality,
meaning by it the insensitivity of the properties of a region of the 
system to perturbations sufficiently far away from it \cite{Kohn96}.
A local language will thus be needed for the two different problems
one has to deal with in a DFT-like method: building the self-consistent
Hamiltonian, and solving it. Most of the initial effort was dedicated 
to the latter \cite{Ordejon95,review} using empirical or semi-empirical 
Hamiltonians. The {\sc Siesta} project \cite{rc,mrs,ijqc} 
started in 1995 to address the former. Atomic-orbital basis sets 
were chosen as the local language, allowing for arbitrary basis sizes, 
what resulted in a general-purpose, flexible linear-scaling DFT program 
\cite{ijqc,prbprep}. A parallel effort has been
the search for orbital bases that would meet the standards of
precision of conventional first-principles calculations, but keeping
as small a range as possible for maximum efficiency.
Several techniques are presented here.

Other approaches pursued by other groups are also 
shortly reviewed in section II. All of them are based on local bases
with different flavors, offering a fair variety of choice between 
systematicity and efficiency. Our developments of atomic bases for 
linear-scaling are presented in section III. 
{\sc Siesta} has been applied to quite varied systems during these years,
ranging from metal nanostructures to biomolecules. Some of the 
results obtained are briefly reviewed in section IV.

\section{METHOD AND CONTEXT}

{\sc Siesta} is based on 
DFT, using local-density \cite{rmp} and generalized-gradients functionals 
\cite{pbe}, including spin polarization, collinear and non-collinear 
\cite{noncol}. 
Core electrons are replaced by norm-conserving pseudopotentials \cite{tm2} 
factorized in the Kleinman-Bylander form \cite{kb}, including
scalar-relativistic effects, and non-linear partial-core corrections 
\cite{pcec}. The one-particle problem is then solved using linear combination
of atomic orbitals (LCAO). There are no constraints either on the
radial shape of these orbitals (numerically treated), or
on the size of the basis, allowing for the full
quantum-chemistry know-how \cite{huzinaga} (multiple-$\zeta$,
polarization, off-site, contracted, and diffuse orbitals).
Forces on the atoms and the stress tensor are obtained from the 
Hellmann-Feynman theorem with Pulay corrections \cite{rc}, and are used for
structure relaxations or molecular dynamics simulations of different
types.

Firstly, given a Hamiltonian, the
one-particle Schr\"odinger equation is solved yielding the energy
and density matrix for the ground state. This task is performed either
by diagonalization (cube-scaling, appropriate for systems under a hundred
atoms or for metals) or with a linear-scaling algorithm. 
These have been extensively reviewed elsewhere \cite{review}. 
{\sc Siesta} implements two $O(N)$ algorithms 
\cite{Ordejon95,kim} based on localized Wannier-like
wavefunctions.

Secondly, given the density matrix, a new Hamiltonian matrix is
obtained. There are different ways proposed in the literature to perform
this calculation in order-$N$ operations.

$(i)$ Quantum chemists have explored algorithms for Gaussian-type orbitals 
(GTO) and related technology \cite{huzinaga}. The long-range Hartree potential 
posed an important problem that has been overcome with Fast Multipole 
Expansion techniques plus near-field corrections \cite{head}. 
Within this approach, periodic boundary conditions for extended systems 
require additional techniques that are under current development \cite{pbcscu}.

$(ii)$ Among physicists tradition favors
more systematic basis sets, such as plane-waves and variations thereof.
Working directly on a real-space grid was early proposed as
a natural possibility for linear scaling \cite{Hernandez95}. 
Multigrid techniques allow efficient
treatment of the Hartree problem, making it very attractive. However,
a large prefactor was found \cite{Hernandez95} for the linear 
scaling, making the order-$N$ calculations along this line not so practical 
for the moment. The introduction of a basis of localized functions on the 
points of the grid (blips) was then proposed as an operative method within 
the original spirit \cite{Hernandez97}. It is probably more expensive 
than LCAO alternatives, but with the advantage of a systematic basis.
Another approach \cite{haynes} works with spherical Bessel
functions confined to (overlapping) spheres wisely located within
the simulation cell. As for plane-waves, a kinetic energy cutoff 
defines the quality of the basis within one sphere. The number, 
positioning, and radii of the spheres are new variables to consider,
but the basis is still more systematic than within LCAO.

$(iii)$ There are mixed schemes that use atomic-orbital bases but evaluate 
the matrix elements using plane-wave or real-space-grid techniques. 
The method of Lippert {\it et al.} \cite{hutter1} uses GTO's and
associated techniques for the computation of the matrix
elements of some terms of the Kohn-Sham Hamiltonian. It uses
plane-wave representations of the density for the calculation of the 
remaining terms. This latter method is conceptually very similar 
to the one presented earlier by Ordej\'on {\it et al.} \cite{rc}, 
on which {\sc Siesta} is based. 
The matrix elements within {\sc Siesta} are also calculated in two
different ways \cite{ijqc}: some Hamiltonian terms in a real-space grid
and other terms (involving two-center integration) by very efficient,
direct LCAO integration \cite{sankey}. While {\sc Siesta} uses numerical 
orbitals, Lippert's method works with GTOs, which allow analytic
integrations, but require more orbitals.

Except for the quantum-chemical approaches, the methods mentioned
require smooth densities, and thus soft pseudopotentials. A recent 
augmentation proposal \cite{hutter2} allows a substantial improvement in
grid convergence of the method of Lippert {\it et al.} \cite{hutter1}, 
possibly allowing for all-electron calculations.

\section{ATOMIC ORBITALS ADAPTED TO LINEAR SCALING}

The main advantage of atomic orbitals is their efficiency (fewer orbitals 
needed per electron for similar precision) 
and their main disadvantage is the lack of systematics for optimal 
convergence, an issue that quantum chemists have been working on for
many years \cite{huzinaga}. They have also clearly shown that there
is no limitation on precision intrinsic to LCAO.

{\it Orbital range.}
The need for locality in linear-scaling algorithms imposes 
a finite range for matrix elements, which has a strong influence on the 
efficiency of the method. There is a clear challenge ahead for finding
short-range bases that still give a high precision.
The traditional way is to neglect matrix elements 
between far-away orbitals with values below a tolerance.
This procedure implies a departure from the original Hilbert space
and it is numerically unstable for short ranges. Instead, the use 
of orbitals that would strictly vanish beyond a certain radius was proposed
\cite{sankey}. This gives sparse matrices consistently within the
Hilbert space spanned by the basis, numerically robust even for small
ranges. 

In the context of {\sc Siesta}, the use of pseudopotentials imposes 
basis orbitals adapted to them. Pseudoatomic orbitals (PAOs) are 
used, i.e., the DFT solution of the atom with the pseudopotential.
PAO's confined by a spherical infinite-potential wall \cite{sankey}, 
has been the starting point for our bases.
Fig.~1 shows $s$ and $p$ confined PAOs for oxygen.
Smoother confining potentials have been proposed as a better converging
alternative \cite{horsfield}.

A single parameter that defines the confinement radii of different 
orbitals is the orbital {\it energy shift} \cite{jose}, 
$\Delta E_{\small \rm PAO}$, i.e., the energy increase 
that each orbital experiences when confined to a finite sphere. It
defines all radii in a well balanced way, and allows the systematic 
convergence of physical quantities to the required precision. 
Fig.~2 shows the convergence of geometry and cohesive 
energy with $\Delta E_{\small \rm PAO}$ for various
systems. It varies depending on the system and physical quantity, but
$\Delta E_{\small \rm PAO} \approx 100 - 200$ meV gives 
typical precisions within the accuracy of current GGA functionals.
%

{\it Multiple-$\zeta$.}
To generate confined multiple-$\zeta$ bases,
a first proposal \cite{projection} suggested the use of the 
excited PAOs in the confined atom. It works well for short ranges, 
but shows a poor convergence with $\Delta E_{\small \rm PAO}$, 
since some of these orbitals are unbound in the free atom.
In the split-valence scheme, widely used in quantum chemistry,
GTOs that describe the tail of the atomic orbitals are
left free as separate orbitals for the extended basis.
Adding the quantum-chemistry \cite{huzinaga} GTOs' tails
to the PAO bases gives flexible bases, but the confinement control 
with $\Delta E_{\small \rm PAO}$ is lost. 
The best scheme used in {\sc Siesta} calculations so far is based
on the idea \cite{joseluis} of adding, instead of a GTO, a numerical
orbital that reproduces the tail of the PAO outside a radius 
$R_{\rm DZ}$, and continues smoothly towards the origin as $r^l(a-br^2)$,
with $a$ and $b$ ensuring continuity and differenciability at $R_{\rm DZ}$.
This radius is chosen so that the norm of the tail beyond has a given
value. Variational optimization
of this {\it split norm} performed on different systems 
shows a very general and stable performance for values around
15\% (except for the $\sim 50\%$ for hydrogen). Within exactly the same 
Hilbert space, the second orbital can be chosen as the difference between
the smooth one and the original PAO, which gives a basis orbital strictly
confined within the matching radius $R_{\rm DZ}$, i.e., smaller than the
original PAO. This is illustrated in Fig.~1. Multiple-$\zeta$ is
obtained by repetition of this procedure.

{\it Polarization orbitals.}
A shell with angular momentum $l+1$ (or more shells with higher $l$) 
is usually 
added to polarize the most extended atomic valence orbitals ($l$), giving
angular freedom to the valence electrons. The (empty) $l+1$ atomic orbitals
are not necessarily a good choice, since they are typically too extended. 
The normal procedure within quantum chemistry \cite{huzinaga} is using
GTOs with maximum overlap with valence orbitals.
Instead, we use for {\sc Siesta} the numerical orbitals resulting from the 
actual polarization of the pseudoatom in the presence of a small electric
field \cite{jose}. The pseudoatomic problem is then exactly solved 
(within DFT), yielding the $l+1$ orbitals through comparison with 
first order perturbation theory. The range of the polarization
orbitals is defined by the range of the orbitals they polarize.
It is illustrated in Fig.~3 for the $d$ orbitals
of silicon.

The performance of the schemes presented here has been tested for various 
applications (see below) and a systematic study will be presented elsewhere
\cite{prbprep}. It has been found in general that double-$\zeta$, singly
polarized (DZP) bases give precisions within the accuracy of GGA functionals
for geometries, energetics and elastic/vibrational properties.

{\it Other possibilities.}
Scale factors on orbitals are also used, both for orbital contraction and for 
diffuse orbitals. Off-site orbitals can be introduced. They serve for
the evaluation of basis-set superposition errors 
\cite{maider}. Spherical Bessel functions are also included, 
that can be used for mixed bases between our approach 
and the one of Haynes and Payne \cite{haynes}.

\section{BRIEF REVIEW OF APPLICATIONS}

{\it Carbon Nanostructures.}
A preliminary version of {\sc Siesta} was first applied to study the
shape of large hollow carbon fullerenes \cite{rc} up to C$_{540}$, 
the results contributing to establish
that they do not tend to a spherical-shape
limit but tend to facet around the twelve corners given by the pentagons.
{\sc Siesta} has been also applied to carbon nanotubes. In a first study,
structural, elastic and vibrational properties were characterized
\cite{tubephonons}. A second work was dedicated to their 
deposition on gold surfaces, and the STM images that they originate
\cite{stmprl}, specially addressing experiments on finite-length tubes.
A third study has been dedicated to the opening of single-wall nanotubes
with oxygen, and the stability of the open, oxidized tubes for intercalation
studies \cite{marioprl}.

{\it Gold Nanostructures.}
Gold nanoclusters of small sizes (up to Au$_{75}$) were found \cite{auprl}
to be amorphous, or nearly so, even for sizes for which very favorable 
geometric structures had been proposed before. In a further study the
origin of this striking situation is explained in terms of local stresses
\cite{auprep}.
Chains of gold atoms have been studied addressing the experiments 
\cite{auexp} which show them displaying remarkably long interatomic spacings 
(4 - 5 \AA). A first study \cite{tosatti} arrives at the conclusion that a 
linear gold chain would break at interatomic spacings much smaller than the
observed ones. It is illustrated in Fig.~4 \cite{auhilos}. A possible
explanation of the discrepancy is reported elsewhere.\cite{auhilos}

{\it Surfaces and Adsorption.}
A molecular dynamics simulation was performed \cite{gabriel} on the clean 
surface of liquid silicon close to the melting temperature, in which
surface layering was found, i.e., density oscillations of roughly atomic
amplitude, like what was recently found to happen in the surface of other
liquid metals \cite{layering}. Unlike them, though, the origin for silicon
was found to be orientational, reminescent of directed octahedral bonding.
Adsorption studies have also been performed on solid silicon surfaces,
Ba on Si(100) \cite{basi} and C$_{60}$ on Si(111) \cite{c60si}. 
Both works study adsorption geometries and energetics. For Ba, interactions
among adsorbed atoms and diffusion features are studied. For C$_{60}$,
STM images have been simulated and compared to experiments.

{\it Nucleic Acids.}
Feasibility tests on DNA were performed in the early stages of the 
project, by relaxing a dry B-form poly(dC)-poly(dG) structure with a 
minimal basis \cite{mrs,ijqc}. In preparation of realistic calculations,
a thorough study \cite{maider} of 30 nucleic acid pairs has been 
performed addressing the precision of the approximations 
and the DZP bases, and the accuracy of the GGA functional \cite{pbe},
obtaining good results even for the hydrogen bridges.
Based on that, a first study of dry A-DNA
has been performed, with a full relaxation of the structure, and an
analysis of the electronic characteristics \cite{dnaprep}.

\section{CONCLUSIONS}

The status of the {\sc Siesta} project has been briefly reviewed,
putting it in context with other methods of liner-scaling DFT, and
briefly describing results obtained with {\sc Siesta} for a variety of
systems. The efforts dedicated to finding schemes for atomic bases adapted
to linear-scaling have been also described. A promising
field still very open for future research.

\vspace{10pt}

{\it Acknowledgments.}
We are grateful for ideas, discussions, and support of
Jos\'e L. Martins, Richard M. Martin, David A. Drabold, Otto F. Sankey,
Julian D. Gale, and Volker Heine. EA is very grateful to the Ecole 
Normale Sup\'erieure de Lyon for its hospitality.
PO is the recipient of a Sponsored Research Project from
Motorola PCRL. EA and PO acknowledge travel support of the $\Psi_k$
network of ESF.
This work has been supported by Spain's DGES grant PB95-0202.

\begin{figure}
\caption[dan1]{Confined pseudoatomic orbitals for oxygen. $s$ in (a) and (b).
$p$ in (c) and (d). $R_c$ is the confinement radius obtained for
$\Delta E_{\small \rm PAO} = 250$ meV. The original PAOs are represented with 
thinner lines. The split smooth functions 
are plotted with thicker lines in (a) and (c), while the resulting
double-$\zeta$ orbitals are plotted with thicker lines in (b) and (d).}
\end{figure}

\begin{figure}
\caption[ea]{Convergence with energy shift $\Delta E_{\small \rm PAO}$ 
of (a) lattice parameters of bulk Si ($\circ$), Au ($\star$), and MgO
($\bullet$), and bond length ($\triangle$) and angle ($\times$)
of H$_2$O; and (b) corresponding cohesive (bond) energies.}
\end{figure}

\begin{figure}
\caption[dan2]{$d$ polarization orbitals for silicon for two different
confinement conditions. (a) Obtained with the electric-field polarization 
method, and (b) the confined $d$ PAOs.} 
\end{figure}

\begin{figure}
\caption[au]{Cohesive energy (a), and stretching force (b) in a linear
gold chain as a function of interatomic distance. Black dots are for
the translationally invariant chain, white circles and squares are for
supercells of 4 and 8 atoms, respectively, where the system is allowed
to break.}
\end{figure}

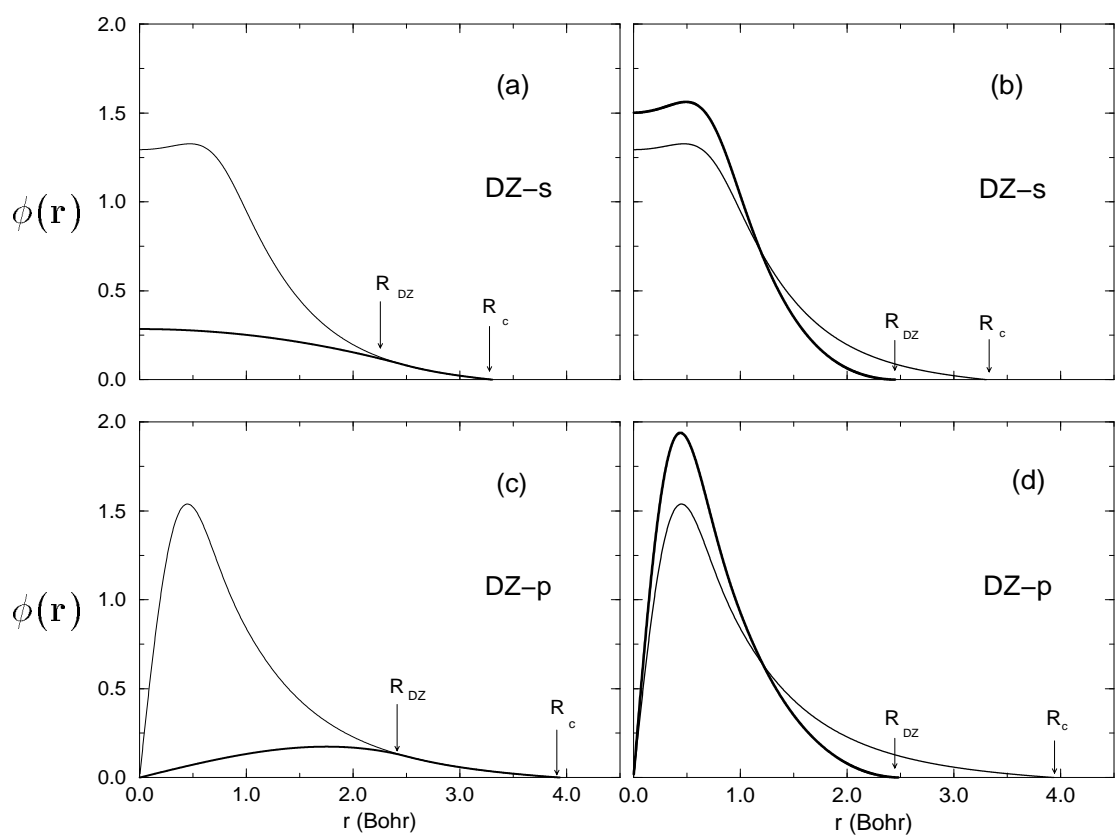
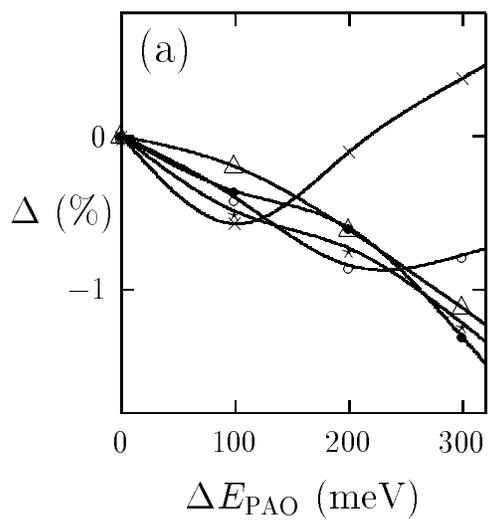
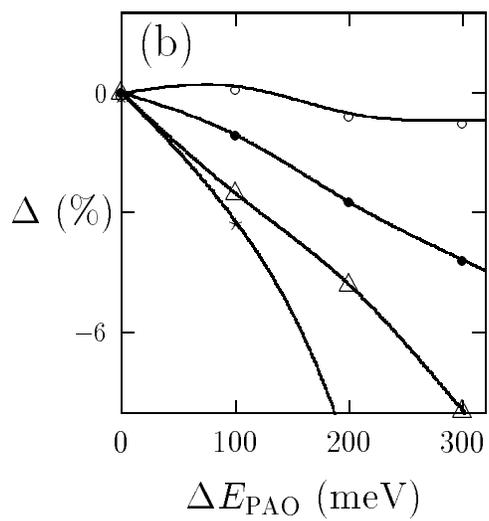
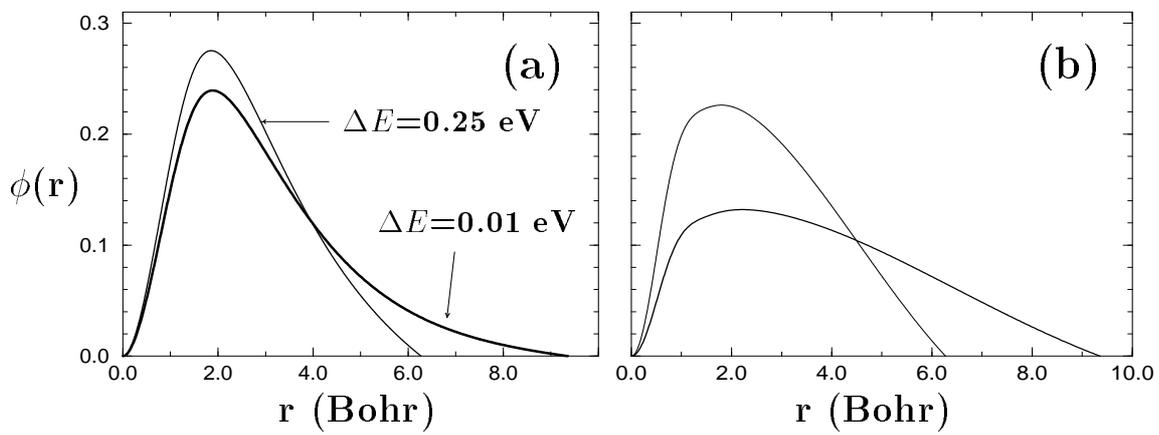
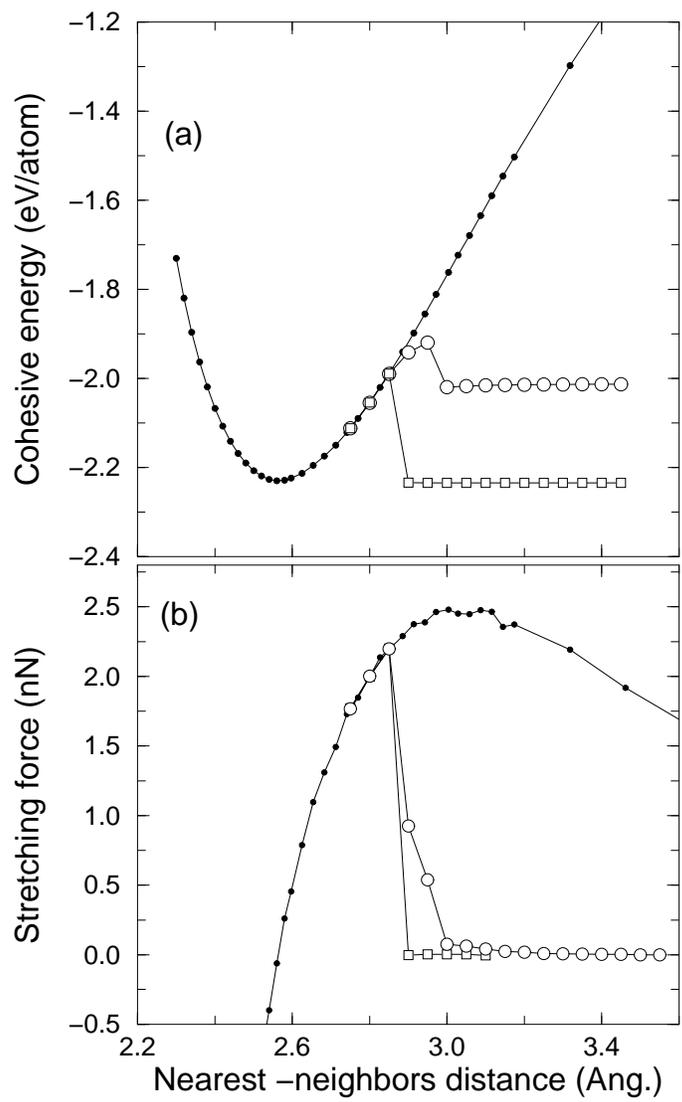
\end{document}